\documentclass{article} 
\usepackage{styles,times}
\usepackage{lipsum}
\usepackage{multirow}
\usepackage{booktabs}
\usepackage{multirow}
\usepackage{multicol}
\usepackage{adjustbox}
\usepackage{tcolorbox}
\usepackage{tabularx}
\usepackage{tikz}
\usepackage{graphicx}
\usepackage{listings}

\usepackage{amsmath,amsfonts,bm}

\def\eqref#1{equation~\ref{#1}}

\def\1{\bm{1}}

\DeclareMathAlphabet{\mathsfit}{\encodingdefault}{\sfdefault}{m}{sl}
\SetMathAlphabet{\mathsfit}{bold}{\encodingdefault}{\sfdefault}{bx}{n}

\usepackage{hyperref}
\usepackage{url}
\usepackage{lineno}
\usepackage{array, makecell}
\usepackage{algorithm}
\usepackage{algpseudocode}
\usepackage[numbers,authoryear]{natbib}
\usepackage{geometry}
\usepackage{amssymb}
\usepackage{chngcntr}
\usepackage{subcaption}
\usepackage{enumitem}
\usepackage{overpic}

\usepackage{float} 
\lstdefinestyle{mcpprompt}{
    basicstyle=\ttfamily\scriptsize,
    breaklines=true,
    breakatwhitespace=true,
    columns=fullflexible,
    keepspaces=true,
    frame=single,
    framesep=4pt,
    xleftmargin=8pt,
    xrightmargin=8pt,
    aboveskip=6pt,
    belowskip=6pt,
    showstringspaces=false,
}
\title{\textsc{LATHE: LAnguage-driven Toolkit for Hypothesis-based crystal Editing}}

\author{Qianyu Zheng$^{\dag}$, Shuyi Jia$^{\dag\text{\textsection}}$, Victor Fung$^{\dag\text{\textsection}}$ \\
{\normalsize $^\dag$College of Computing, Georgia Institute of Technology, Atlanta, GA, USA} \\
{\normalsize $^\ddag$School of Computational Science and Engineering, Georgia Institute of Technology, GA, USA}\\
{\normalsize $^{\text{\textsection}}$Corresponding authors: 
\texttt{victorfung@gatech.edu}}
}
\date{}

\newcolumntype{Y}{>{\centering\arraybackslash}X}

\iclrfinalcopy 
\begin{document}

\maketitle

\begin{abstract}

Manipulating crystal structures towards targeted geometric properties or symmetry constraints is a longstanding challenge. Existing computational frameworks fall short: LLMs have poor spatial awareness and operate at coarse granularity in the token space, while diffusion- and gradient-based structure generation approaches generally produce complete structures \textit{de novo} rather allowing for fine-grained editing of a given input under arbitrary constraints. We present LATHE, a geometric toolkit that closes this gap by expressing seven classes of crystallographic properties --- bond length, bond angle, dihedral angle, coordination environment, lattice parameters, cell volume, and space group, as differentiable objectives to enable direct geometric editing of crystal structures. To demonstrate its usage in materials design, we further expose LATHE through a Model Context Protocol server and embed it in a closed-loop multi-agent system which translates hypotheses intogeometric modifications in natural language towards a given design objective. Across single-property benchmarks, LATHE attains near-perfect constraint satisfaction on all seven property types while keeping optimized structures close to local energy minima. The LATHE-equipped agent translates over 87.5\% of natural-language prompts into valid executable configurations and faithfully completes them. In a band-gap inverse-design case study, the multi-agent loop reaches the target tolerance window in nine of ten independent runs at a typical cost of thirteen hypothesis-evaluation cycles. By bridging natural-language hypothesis generation and physically grounded gradient-based structural editing, this work establishes a paradigm for interpretable, closed-loop computational materials discovery that is immediately applicable to a broad class of functional material design tasks.
\end{abstract}

\section{Introduction}\label{sec:intro}

Designing materials with targeted functional properties---such as a prescribed band gap, specific coordination geometry, or desired crystal symmetry---remains a longstanding challenge in materials science due to the combinatorial vastness of composition and structure spaces (\cite{Zhang_2025, Curtarolo2013-ts, Lookman2019-nk}). Traditional approaches to materials discovery often rely on high-throughput computational screening via density functional theory (DFT), which isresource-intensive and inefficient compared to iterative, hypothesis-driven workflows (\cite{Jain2011-jb, Jain2013-iq}). Meanwhile, recent advances in deep generative models have introduced an array of different generative frameworks for conditional generation of crystal structures. These approaches include foundational equivariant diffusion models operating directly in atomic coordinate space (\cite{xie2022crystaldiffusionvariationalautoencoder, jiao2024crystalstructurepredictionjoint, Zeni2025-qn}), autoregressive language models that treat crystal structure files as text sequences (\cite{Antunes2024-ph}), flow-matching and Riemannian manifold-based generative frameworks (\cite{miller2024flowmmgeneratingmaterialsriemannian, tangsongcharoen2025crystalgrwgenerativemodelingcrystal, sinha2024representationspacediffusionmodelsgenerating}), and property- or constraint-guided approaches that steer generation toward targeted electronic structures or geometric motifs (\cite{jia2025electronicstructureguidedinverse, Okabe2026-vf}). Despite their differing strategies and advantages, these approaches share a common limitation: they operate at the level of generation or end-to-end property targeting and do not provide a controllable, physically grounded mechanism for modifying an existing structure in response to a specific geometric hypothesis. In computational materials discovery, a common practice is to investigate how modifications of certain sites or motifs of structures (e.g. bond elongation, local environment tweaks or symmetry changes) impact the target property. Yet, when a materials scientist hopes to investigate impacts of such fine-grained structural modifications, no existing computational framework can answer that question by faithfully presenting the structure with the desired changes. A different paradigm is therefore needed for material discovery to bridge qualitative structural intuition and quantitative geometric control at the atomic scale.

In recent years there have been a dramatic expansion in the capabilities of large language models (LLMs), driven by ever-growing training corpus and unprecedented increases in computing scale (\cite{brown2020languagemodelsfewshotlearners, kaplan2020scalinglawsneurallanguage, hoffmann2022trainingcomputeoptimallargelanguage}). Beyond text generation, LLMs have demonstrated the ability to reason, reflect on prior decisions (\cite{wei2023chainofthoughtpromptingelicitsreasoning}), and act as autonomous agents when augmented with external tools (\cite{yao2023react, schick2023toolformerlanguagemodelsteach, mialon2023augmentedlanguagemodelssurvey}). These capabilities have attracted growing interest in scientific domains, particularly materials science and chemistry (\cite{Lei2024-wy, Jablonka2023, Ramos2025}). Prior work has demonstrated that LLM-based multi-agent systems can propose compositional modifications to drive target material properties: \cite{bran2024chemcrow} and \cite{Boiko2023-wk} integrate LLMs with external chemistry tools and experimental automation to autonomously plan and execute chemical syntheses and guide molecular discovery; \cite{jia2024llmatdesign} and \cite{ansari2024dzinerrationalinversedesign} employ self-reflective LLM agents to iteratively propose and evaluate compositional modifications for property-driven inverse design; \cite{ghafarollahi2024atomagentsalloydesigndiscovery} and \cite{ghafarollahi2024rapidautomatedalloydesign} couple physics-aware multi-agent systems with atomistic simulations and graph neural networks to autonomously navigate alloy composition spaces and predict mechanical properties; and \cite{zhang2024honeycombflexiblellmbasedagent}, \cite{Takahara_2025} and \cite{Chaudhari2026-ty} build modular agent frameworks for broader materials science workflows, pairing LLM reasoning with curated knowledge bases, diffusion-based structure generation, and domain-specific simulation tools to direct exploration toward targeted material properties. Yet, these frameworks mostly operate at the coarse granularity of elemental substitution, simple site displacement or structure modification with attached code execution tools, and have difficulty faithfully completing precise, continuous geometric operations, such as bond stretching, angle tilting, coordination shell reshaping, symmetry enforcement, which are central to structure–property reasoning. Crucially, there lacks a back-end capable of executing such operations faithfully, leaving a significant gap between what an LLM agent can hypothesize in natural language and what it can actually manipulate in the crystal structure space.

In this work, we present a \textbf{LA}nguage-driven \textbf{T}oolkit for \textbf{H}ypothesis-based crystal \textbf{E}diting (LATHE) that closes this gap between the proposal and the implementation of the geometric hypotheses. Concretely, LATHE expresses seven classes of geometric properties --- bond length, bond angle, dihedral angle, coordination environment, lattice parameters, cell volume, and space group --- as differentiable loss functions over fractional coordinates and the lattice parameters to enable direct gradient-based optimization of any combination of these properties. An auxiliary machine-learning interatomic potential (MLIP) acts as a differentiable energy regularizer to keep optimized structures physically plausible without overriding the geometric objective. We expose LATHE through a Model Context Protocol (MCP) server (\cite{mcp_spec_2025}), providing resource and tool endpoints that allow an LLM agent to translate natural-language input instructions into validated optimization configurations and execute them accordingly. Building on this agent, we demonstrate a multi-agent system in the spirit of LLMatDesign (\cite{jia2024llmatdesign}), in which a Hypothesis Agent proposes geometric modifications in natural language and the LATHE-equipped Agent translates each proposal into a configuration for gradient-based optimization; the resulting structure is evaluated for the target property, and the (hypothesis, modification, outcome) triple is fed back to close the hypothesize--translate--optimize--evaluate loop.

Our framework differs from prior gradient-based and generative approaches in three respects. First, whereas the most relevant work, SMOACS (\cite{fujii2025rethinkinggradientbasedmethodsmultiproperty}), optimizes through learned property predictors with template-derived initial conditions, our losses are explicit functions of the fractional coordinates and lattice matrix, exposing each property as an interpretable target rather than a single black-box objective. Second, LATHE is built around \emph{editing an existing structure} under a user- or hypothesis-specified geometric constraint rather than synthesizing a representative of the target space group from scratch as seen in constraint-conditioned generative methods (\cite{Okabe2026-vf, sinha2024representationspacediffusionmodelsgenerating}), which makes it suitable for being integrated into iterative crystal structure design pipelines. Third, LATHE can be wrapped in a language-accessible agent layer that translates qualitative chemical hypotheses into executable optimization configurations within a closed feedback loop, an integration absent from prior gradient-based crystal optimization work. The result is an interpretable-by-construction design trajectory in which every accepted structural change carries a human-readable hypothesis and a measured property outcome. We demonstrate that LATHE achieves near-perfect constraint satisfaction across all seven property types in single-objective optimization, that the LATHE-equipped Agent correctly translates and carries out over 87.5\% of natural-language prompts, including deliberately ambiguous ones, into valid executable configurations, and that the multi-agent loop reliably drives a target band gap to within a user-specified tolerance.

\section{Methodology}\label{sec:method}

\subsection{Graph-Based Crystal Structure Representation}

Crystal structures are represented as attributed graphs $\mathcal{G} = (\mathcal{V}, \mathcal{E})$, where nodes $\mathcal{V}$ correspond to atoms and edges $\mathcal{E}$ encode interatomic interactions within a periodic unit cell. Geometric property optimization is formulated as a continuous optimization problem over two differentiable tensors: the atomic fractional coordinates $\mathbf{F} \in \mathbb{R}^{N \times 3}$ and the lattice matrix $\mathbf{L} \in \mathbb{R}^{3 \times 3}$, whose rows are the lattice vectors. Cartesian positions are recovered as $\mathbf{X} = \mathbf{F} \cdot \mathbf{L}$, and all geometric properties are computed from $\mathbf{X}$ and $\mathbf{L}$, preserving differentiability through both tensors. For geometric property losses considered in this project, including bond length, bond angle, dihedral, coordination environment, lattice parameters, cell volume, and space group, edges are constructed using the minimum image convention (MIC), selecting for each atom its nearest periodic images within a fixed cutoff. For energy evaluation, MatterSim (\cite{yang2024mattersim}) is used as the machine learning interatomic potential (MLIP), and the graph with the two-body and three-body interaction indices (with PBC considered) is constructed to exactly match MatterSim's native input convention. To avoid the computational overhead and inconsistent gradient signals between optimization steps due to rebuilding the neighbor list at every optimization step, we adopt a Verlet-list caching strategy (\cite{Verlet-list}): the discrete graph topology is rebuilt only when the maximum Cartesian displacement of any atom from its position at the last rebuild exceeds a threshold of 0.25 Å, while the differentiable quantities $\mathbf{X}$ and $\mathbf{L}$ are recomputed with gradients at every step.

\begin{figure}
    \centering
    \includegraphics[width=1\linewidth]{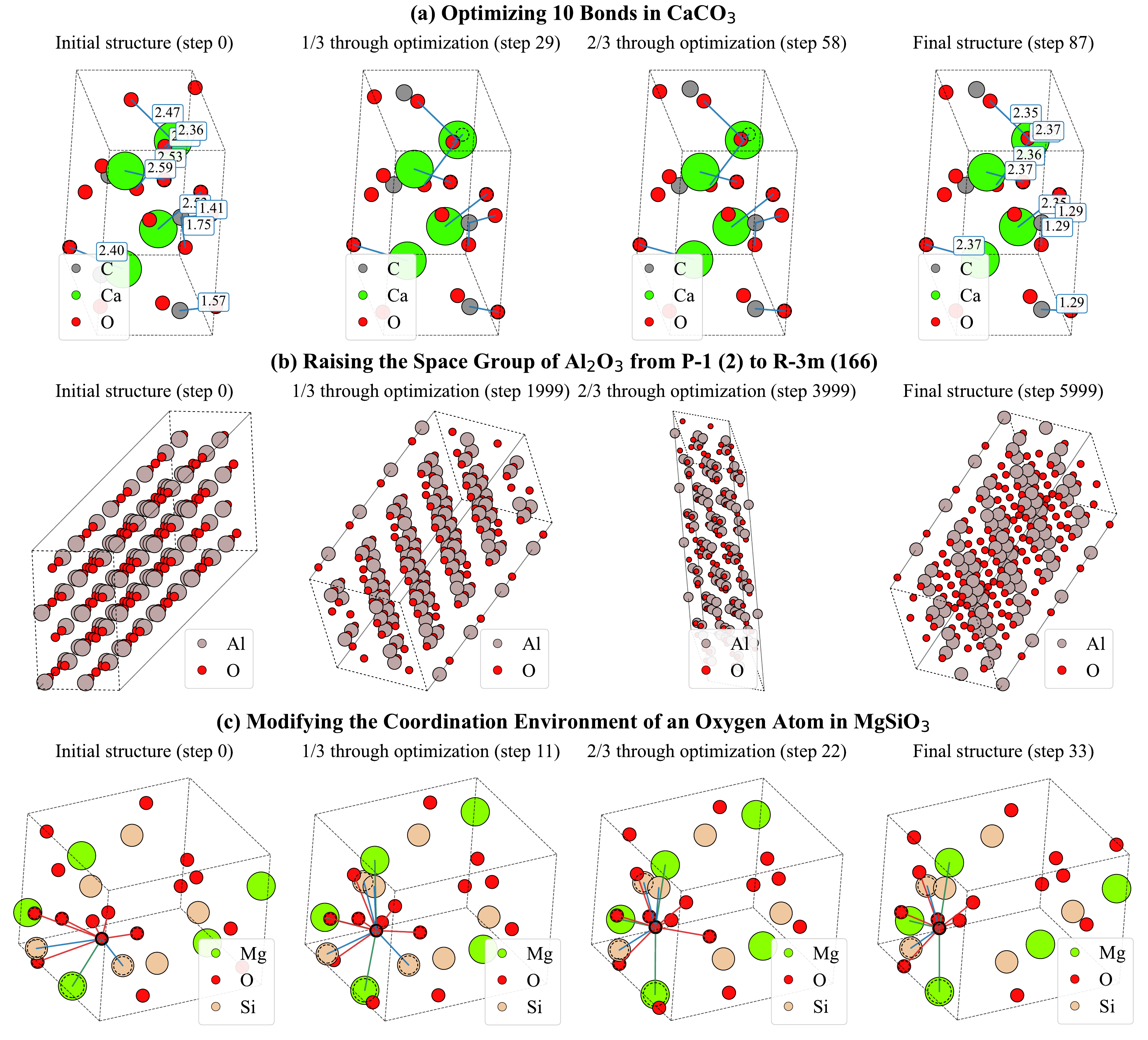}
    \caption{\textbf{Representative gradient-optimization trajectories.} Each row shows four frames of a single optimization — the initial structure, frames at approximately 1/3 and 2/3 of the trajectory, and the final optimized structure, with the step index given in parentheses. In every panel, periodic-image partners that lie outside the primitive cell are indicated by a dashed black ring drawn around the atom. \textbf{(a) Bond-length optimization of a CaCO$_3$ structure (mp-556235).} Ten bonds are jointly driven to user-specified distances under a weak MLIP-energy co-objective ($w = 0.01$). Each targeted $(i, j)$ pair is drawn as a uniform blue line connecting the two atoms (or atom $i$ and the nearest periodic image of atom $j$ when the bond wraps through the cell boundary). The first frame labels each bond with its starting distance (in \AA); the final frame labels each bond with its target distance (in \AA); intermediate frames show the bonds without numerical labels for clarity. \textbf{(b) Space-group elevation of Al$_2$O$_3$ (mp-638765) from P-1 (No. 2) to R-3m (No. 166).} The $2 \times 2 \times 2$ supercell of the structures are shown. The optimization is again regularized by the MLIP-energy co-objective. \textbf{(c) Coordination-environment modification of an oxygen site in MgSiO$_3$ (mp-1180468).} The goal is to raise the number of Mg neighbors of an Oxygen atom from one to three while maintaining the number of Si and O neighbors. The target oxygen atom is highlighted with a bold solid black ring, and its first-neighbor shell within the coordination-loss cutoff of 3.5 \AA~is drawn as colored lines from the target site to each neighbor, one line per (atom, periodic-image) contact and colored by the element of the neighbor (Mg in green, Si in blue; O in red). As the optimization progresses, the number and identity of neighbors inside the cutoff shell evolve toward the specified target coordination.}
    \label{fig:demo_optimization}
\end{figure}

\subsection{Geometric Properties as Differentiable Functions}

In this section, we describe the design of loss functions for each geometric property of interest in LATHE:

\paragraph{Bond length, bond angle, and dihedral angle} The losses for these three properties are computed directly from atomic positions using the minimum image convention (MIC). For a pair of sites $(i, j)$, the MIC displacement vector is obtained by computing the fractional coordinate difference $\Delta\mathbf{f} = \mathbf{f}_j - \mathbf{f}_i$, rounding to the nearest integer image as $\Delta\mathbf{f} \leftarrow \Delta\mathbf{f} - \lfloor \Delta\mathbf{f} \rceil$ to select the shortest periodic image, and projecting into Cartesian space via $\mathbf{v}_{ij} = \Delta\mathbf{f} \cdot \mathbf{L}$, where $\mathbf{L}$ is the lattice matrix. Gradients are propagated through the fractional difference and the lattice projection, while the rounding operation is detached from the computational graph to avoid ill-defined subgradients. The bond length loss is then the squared deviation of $\|\mathbf{v}_{ij}\|$ from a target distance $d^*$. Bond angles at a vertex $j$ in a triplet $(i, j, k)$ are computed from the two MIC vectors $\mathbf{v}_{ji}$ and $\mathbf{v}_{jk}$ using the numerically stable formulation $\theta = \text{atan2}(\|\mathbf{v}_{ji} \times \mathbf{v}_{jk}\|,\, \mathbf{v}_{ji} \cdot \mathbf{v}_{jk})$ (\cite{atan2-bond_angle}), which avoids the gradient singularities of the $\arccos$ formulation near $0$ and $\pi$. Dihedral angles across a quadruplet $(i, j, k, l)$ are computed from the three sequential bond vectors $\mathbf{b}_1 = \mathbf{v}_{ij}$, $\mathbf{b}_2 = \mathbf{v}_{jk}$, $\mathbf{b}_3 = \mathbf{v}_{kl}$ via the signed $\text{atan2}$ formulation over the plane normals $\mathbf{n}_1 = \mathbf{b}_1 \times \mathbf{b}_2$ and $\mathbf{n}_2 = \mathbf{b}_2 \times \mathbf{b}_3$ (\cite{atan2-bond_angle}), which yields a signed angle in $(-\pi, \pi]$. For each property, the loss is the mean squared error (MSE) between the computed value and the user-specified target, averaged over all designated site tuples.

\begin{figure}
    \centering
    \includegraphics[width=1\linewidth]{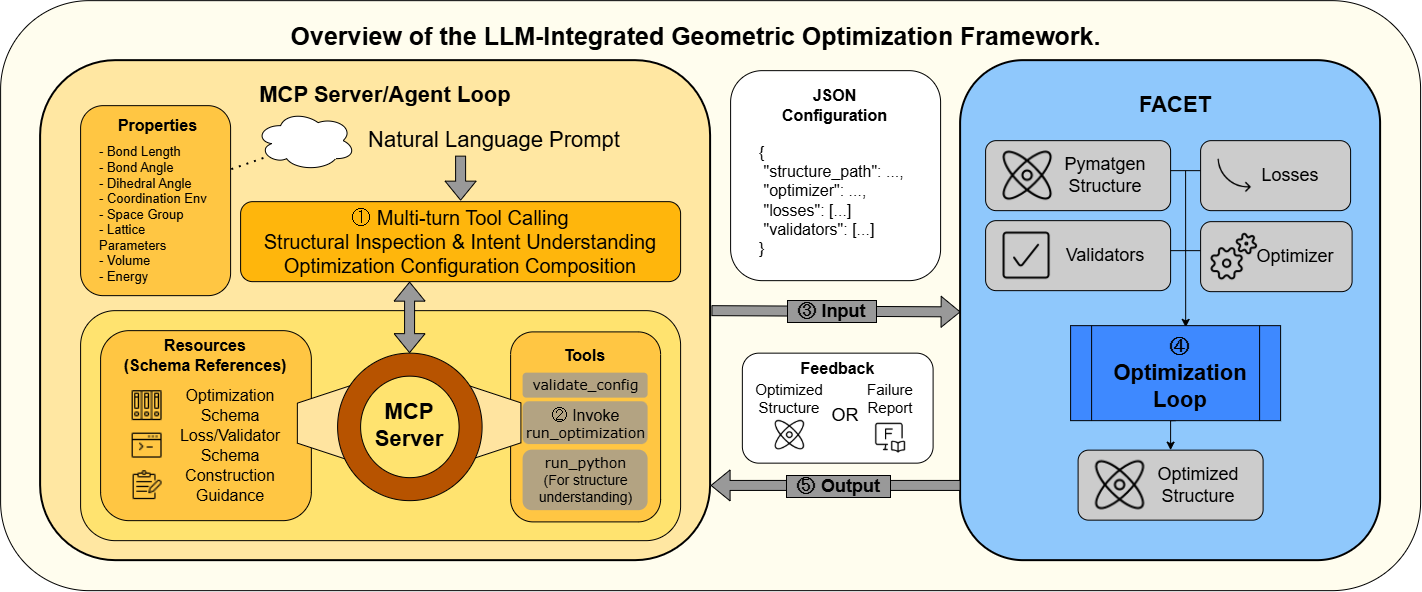}
    \caption{\textbf{Overview of the LATHE-equipped agent.} A natural-language prompt containing properties to optimize is issued to an MCP Server/Agent loop, which iteratively composes a valid optimization configuration through multi-turn tool calling over structural inspection and intent understanding. The agent is grounded by MCP Server resources exposing the configuration, loss, validator, and constraint schemas to compose a JSON configuration defining the optimization targets and specifications, which is dispatched to LATHE via the \texttt{run\_optimization} tool. After receiving the configuration, LATHE loads the structure to optimize, instantiates the user-specified losses and validators, and performs the optimization within a differentiable optimization loop. The optimized structure is returned to the agent to close the feedback cycle between language-level intent and gradient-based crystal structure design.}
    \label{fig:single-agent}
\end{figure}

\paragraph{Coordination Environment} Coordination environment optimization targets the number of neighbors of each species type around a designated central site. Since integer neighbor counts are non-differentiable, we adopt a soft counting scheme based on Gaussian-smeared radial distribution integration (\cite{zimmermann2017, Zimmermann2020yx}), in which each neighboring atom contributes a smooth, distance-dependent weight $c_j \in [0,1]$ to a soft coordination number $\tilde{n} = \sum_j c_j$ (full functional form in SI Section~S1.1). The primary loss term is the squared deviation $(\tilde{n} - n^*)^2$ from the target count $n^*$, summed over all species types. Three auxiliary penalties are added to keep the optimization well-behaved near integer boundaries and to prevent unphysical geometries: a sharpness penalty that discourages ambiguous shell membership, an integer-snapping penalty that removes gradient dead zones at fractional counts, and a soft-core repulsion combined with an attractive well potential that respectively prevent atomic collapse and allow under-coordinated sites to recover across the cutoff boundary. Full functional forms and gating terms are given in SI Section~\ref{sec:SI-coord-env}.

\paragraph{Symmetry} The space group compliance loss $\mathcal{L}_\text{sg}$ is a composite of four differentiable terms acting jointly on the fractional coordinates $\mathbf{F}$ and lattice matrix $\mathbf{L}$:
$$\mathcal{L}_\text{sg} = \mathcal{L}_\text{sym} + \lambda_\text{lat}\,\mathcal{L}_\text{shape} + \lambda_\text{vol}\,\mathcal{L}_\text{vol} + \lambda_\text{floor}\,\mathcal{L}_\text{floor} + \lambda_\text{overlap}\,\mathcal{L}_\text{overlap}$$
\textbf{Symmetry compliance} $\mathcal{L}_\text{sym}$ is the primary term. For each symmetry operation $\{R_g \,|\, \mathbf{t}_g\}$ of the target space group, the image of every atom under that operation should coincide with some atom of the same species. The loss penalizes the mean squared minimum-image Cartesian distance between each symmetry-transformed position $R_g \mathbf{f}_i + \mathbf{t}_g$ and its optimal assignment among atoms of matching species, where the assignment is computed by Hungarian matching (\cite{doi:10.1137/0105003, Kuhn1955-zx}) at each step and treated as a fixed index set for gradient computation. This makes the loss differentiable with respect to $\mathbf{F}$ and $\mathbf{L}$ while handling the combinatorial assignment problem outside the computational graph. \textbf{Lattice shape} $\mathcal{L}_\text{shape}$ penalizes deviations of the lattice metric from the constraints imposed by the target crystal system (e.g., cubic symmetry requires $a=b=c$ and $\alpha=\beta=\gamma=90^\circ$). \textbf{Volume regularization} $\mathcal{L}_\text{vol} = \left(|\det \mathbf{L}| / V_0 - 1\right)^2$ anchors the cell volume to a reference value $V_0$ derived from the input structure, preventing the optimizer from collapsing or inflating the cell to trivially minimize the symmetry term. \textbf{Floor} and \textbf{overlap} penalties apply quadratic soft constraints that prevent degenerate lattice vectors and unphysical atom–atom overlap, respectively.
 
Since space group symmetry is a discrete, global property of the crystal, naive optimization from an arbitrary starting configuration is hard to converge, as documented in our initial experiments. To address this, we initialize atomic positions onto valid Wyckoff sites of the target space group before optimization begins to provide a warm start. This procedure is described in detail in SI Section~\ref{sec:SI-wyckoff-initialization}.

\paragraph{Lattice Parameters and Volume} The lattice parameter loss optimizes the six conventional crystallographic parameters $\{a, b, c, \alpha, \beta, \gamma\}$, three edge lengths in \AA{}, and three interaxial angles in degrees toward user-specified target values. These are computed from the lattice matrix $\mathbf{L}$ as vector norms and pairwise angles between row vectors, preserving differentiability through $\mathbf{L}$. The loss is a weighted sum of squared deviations over lengths and angles separately, $\mathcal{L}_{\text{latt}} = w_{\ell} \sum_{i \in \{a,b,c\}} (p_i - p_i^*)^2 + w_{\alpha} \sum_{i \in \{\alpha,\beta,\gamma\}} (p_i - p_i^*)^2$, where $w_{\ell}$ and $w_{\alpha}$ are independent weights that account for the difference in scale between length and angle targets. Cell volume is computed as the absolute value of the determinant of $\mathbf{L}$, $V = |\det \mathbf{L}|$, and the volume loss is formulated either as an absolute squared deviation $(V - V^*)^2$ or a relative deviation $\left(\frac{V}{V^*} - 1\right)^2$, the latter being preferred when the target volume differs substantially from the initial value to avoid loss scale sensitivity.

\paragraph{Potential Energy} The potential energy can be incorporated as an auxiliary loss term to introduce physical plausibility awareness into the optimization, rather than as a primary objective. Without an energy penalty, gradient-based optimization of isolated geometric properties can drive structures into configurations that satisfy the target constraint but are physically and by extension, energetically unreasonable. To address this, we employ MatterSim-1M (\cite{yang2024mattersim}), a pretrained machine learning interatomic potential (MLIP) trained on a broad distribution of inorganic crystal structures, to evaluate the potential energy of the structure at each optimization step, though any other potential can be used. The predicted energy enters the total loss as a weighted auxiliary term $\mathcal{L}_{\text{total}} = \mathcal{L}_{\text{geo}} + \lambda \mathcal{L}_{\text{energy}}$, where $\mathcal{L}_{\text{geo}}$ is the primary geometric property loss, $\mathcal{L}_{\text{energy}}$ is the MatterSim-predicted potential energy, and $\lambda$ controls the relative contribution of the energy penalty. Since MatterSim is a differentiable graph neural network, gradients flow through it without modification, and the energy loss participates fully in the auto-differentiation graph alongside the geometric losses. The weight $\lambda$ is treated as a user-specified hyperparameter and is kept small relative to the geometric loss scale, ensuring that energy awareness acts as a regularizer that discourages unphysical configurations without overriding the geometric optimization objective.

\subsection{Optimization Algorithms}

\subsubsection{Optimization with Auto-Differentiation}

To optimize crystal structures toward target geometric properties, we treat the fractional coordinates \textbf{F} and the lattice matrix \textbf{L} as parameters and implement two gradient-based strategies: Adam-based gradient descent (\cite{adam}) and Basin-Hopping (\cite{basin-hopping}) with Adam. In both settings, the parameters are updated directly via the Adam optimizer with cosine learning rate scheduling and early stopping criteria for each desired geometric property such that optimization terminates as soon as all target property constraints are simultaneously satisfied, rather than running for a fixed number of iterations. An exception to the early stopping rule is when energy loss is included. To prevent stopping before sufficiently many energy optimization steps are applied, we run the full optimization and pick the candidate structure that satisfies all the geometric properties and has the lowest MLIP-predicted energy. While gradient descent suffices for many property targets, certain objectives, most notably space group conversion, exhibit a landscape with deep, discrete local minima that gradient descent alone cannot escape. In particular, a structure may become trapped in a supergroup of the target space group, where all symmetry operations of the desired space group are satisfied and thus the gradients are too small to break the excess symmetry; gradient signals in such regions are either vanishing or misleading. To address this, we additionally implement Basin-Hopping, in which random perturbations to atomic positions are applied to escape the current basin, with the best-seen structure retained across hops. This two-phase strategy is especially critical for space group optimization, as the finite stochastic jumps allow the structure to break out of supergroup-symmetric configurations and explore lower-symmetry basins that are otherwise inaccessible to gradient-only methods. Examples of optimization trajectory are shown in Figure \ref{fig:demo_optimization}. More comprehensive experimental outcomes are discussed later in Section \ref{sec:single-property-results}.

\subsubsection{Multi-Objective Optimization}

To optimize multiple geometric properties simultaneously, we formulate a composite objective as a weighted sum of individual property losses, where each loss is assigned a scalar weight that controls its relative contribution to the total objective. This allows prioritization of certain structural targets over others, especially over the energy objective as a regularizer. However, naively summing gradients from competing property losses can lead to destructive interference, where gradients from different objectives point in conflicting directions and impede overall convergence. To address this, we adopt Project Conflicting Gradients (PCGrad) (\cite{pcgrad}), which detects pairs of task gradients that have a negative cosine similarity and resolves conflicts by projecting each gradient onto the normal plane of the other to mitigate conflicting gradients from different objectives. This ensures that no single property loss actively undermines the progress of another and enables more stable and balanced convergence across all objectives.
To optimize multiple geometric properties simultaneously, we formulate a composite objective as a weighted sum of individual property losses, where each loss is assigned a scalar weight that controls its relative contribution to the total objective. This allows prioritization of certain structural targets over others, especially over the energy objective as a regularizer. However, naively summing gradients from competing property losses can lead to destructive interference, where gradients from different objectives point in conflicting directions and impede overall convergence. To address this, we adopt Project Conflicting Gradients (PCGrad) (\cite{pcgrad}), which detects pairs of task gradients that have a negative cosine similarity and resolves conflicts by projecting each gradient onto the normal plane of the other to mitigate conflicting gradients from different objectives. This ensures that no single property loss actively undermines the progress of another and enables more stable and balanced convergence across all objectives.

\subsection{Equip LLM Agents with LATHE for LATHE-equipped Skills}

\subsubsection{\label{sec:mcp-server}Integration of LATHE with a MCP Server}

\begin{table}[t]
  \centering
  \begin{tabular}{llp{6.5cm}}
  \toprule
  \textbf{Type} & \textbf{Endpoint} & \textbf{Description} \\
  \midrule
  Resource & \texttt{schema://config}        & Full JSON schema of the optimization configuration. \\
  Resource & \texttt{schema://losses}         & Catalogue of available loss functions with accepted parameters.
  \\
  Resource & \texttt{schema://optimizer}    & Per-optimizer documentation (learning rate schedules, basin-hopping settings, etc.). \\
  \midrule
  Tool     & \texttt{validate\_config}        & Checks a candidate configuration against the Pydantic schema and returns structured diagnostics. \\
  Tool     & \texttt{optimize\_structure}     & Accepts a validated configuration and an input structure, then executes the full optimization pipeline. \\
  Tool     & \texttt{run\_python}             & Executes arbitrary Python code for inspection and analysis inspection of the input crystal structure. \\
  \bottomrule
  
  \end{tabular}
  \caption{Resources and tools exposed by the MCP server to the LATHE-equipped agent.}
  \label{tab:mcp_endpoints}
\end{table}

To expose LATHE as a service consumable by external clients and autonomous agents, we implement a Model Context Protocol (MCP) server (\cite{mcp_spec_2025}) using the FastMCP framework. The server presents capabilities of LATHE through two complementary interface categories: resources and tools. Resources provide read-only access to the system's schema definitions, allowing a client to inspect the design space without side effects, while tools expose actionable endpoints that modify state or trigger computation. Table~\ref{tab:mcp_endpoints} summarizes the full set of endpoints. This resource/tool separation enforces a clean contract between discovery and execution, enabling an LLM agent to first query the available losses, validators, and optimizer settings, then compose a well-formed configuration, and finally submit it for optimization, all through a single HTTP/SSE transport layer without direct access to the underlying Python runtime.

\subsubsection{\label{sec:geom-agent}Design of the LATHE-equipped Agent}

    Following prior works about LLM agents in chemistry/materials (\cite{bran2024chemcrow, Boiko2023-wk, wang2026matty, agada2026review}), we envision the role of the autonomous agent to be translating natural-language crystal-engineering requirements into executable optimization configurations and implemented it as a Pydantic AI agent (\cite{pydanticai2026}) that communicates with the MCP server as its sole back-end, with the complete design implemented as shown in Figure~\ref{fig:single-agent}. This LATHE-equipped agent enforces a strict multi-phase workflow to ensure correctness: it first inspects the input structure by querying its geometric properties, then consults the MCP server's schema resources to identify the appropriate loss functions and validators, and subsequently assembles a candidate configuration for interacting with LATHE. Before execution, the agent performs a quality check to verify that every target value falls within a physically plausible range for the given structure and conducts a validator audit to confirm that each constraint specified in the natural-language prompt is covered by a corresponding validator with matching tolerances. Only after the configuration passes the MCP server's tool for confirming the correctness of the JSON optimization configuration is it submitted to the main \texttt{optimize\_structure} tool for gradient-based refinement. This Chain-of-Thought (\cite{wei2022cot}) and ReAct (\cite{yao2023react}) style multi-stage pipeline, enabled by prompting, prevents the agent from creating under-specified objectives and enables the agent to provide interpretable feedback at each decision point, bridging the gap between high-level materials-design intent and the low-level parameterization required by the differentiable optimization engine.

\subsubsection{\label{sec:multi-agent-design}A Multi-Agent System for LATHE-driven Structural Design}

The LATHE-equipped Agent reliably executes well-specified structural modifications, but realistic materials-design tasks are rarely posed at that level of granularity. A practitioner targeting a desired electronic, mechanical, or thermodynamic property seldom knows \emph{a priori} which combination of bond lengths, coordination environments, or symmetry constraints will yield the intended outcome. To bridge this gap, we integrate the LATHE-equipped Agent into a closed-loop multi-agent system, in which a second large-language-model agent, the \emph{Hypothesis Agent}, reasons about structure-property relationships and proposes successive modifications that drive a target property toward a user-specified value. We use the task of optimizing a structure towards a desired band gap as the example to demonstrate the effectiveness of this system.

\begin{figure}
  \centering
  \includegraphics[width=1\linewidth]{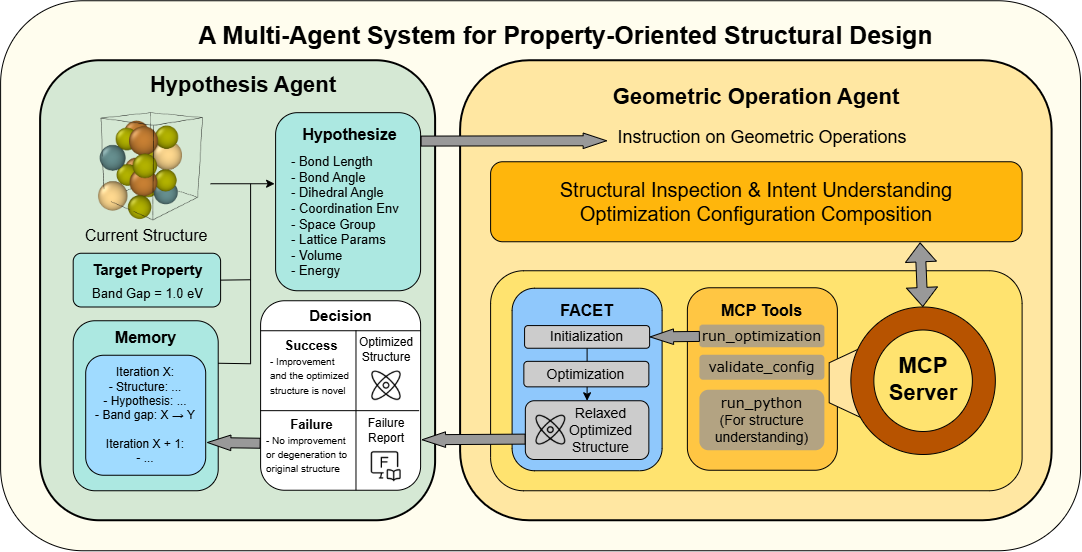}
  \caption{\textbf{A two-agent system for property-oriented structural design.} The framework decomposes inverse design into two cooperating agents. The Hypothesis Agent (left) conditions on the current structure and a user-specified target property to propose candidate geometric hypotheses --- bond lengths, bond and dihedral angles, coordination environments, space group, lattice parameters, volume, and energy --- while maintaining a memory of prior iterations that supports re-hypothesis upon failure and informs subsequent proposals. Each hypothesis is passed as an instruction to the LATHE-equipped Agent (right), which performs structural inspection and intent understanding to compose an optimization configuration.}
  \label{fig:multi-agent}
\end{figure}

Following the design of LLMatDesign (\cite{jia2024llmatdesign}), we adopt in-context learning as the framework for structural design; the system is illustrated in Figure~\ref{fig:multi-agent}. At each iteration, the Hypothesis Agent receives the current crystal structure (in CIF form), the present and target values of the property of interest, the admissible vocabulary of geometric operations exposed by LATHE, and a complete record of previously attempted hypotheses together with their measured property outcomes and any post-relaxation verdicts. Conditioned on this context, it generates a natural-language instruction that is dispatched to the LATHE-equipped Agent for translation and execution, the resulting structure is evaluated for the target property, and the (hypothesis, modification, property) triple is appended to the exploration history that conditions subsequent proposals. This resembles the multi-agent scientific hypothesis testing frameworks well documented in the existing literature (\cite{Gottweis_2026, yamada2025aiscientistv2workshoplevelautomated, yuksel2024multiaiagentautonomousoptimization}).

We evaluate band-gap optimization under two settings of increasing physical constraints: \emph{manipulation} and \emph{design}. The \emph{manipulation} setting is a test of whether the agentic system can simply find a configuration of atomic positions and lattice parameters to achieve the target property (band gap) with LATHE without considering other factors such as whether it is in a local minima. Such purely geometric levers on the band gap --- strain, bond-length and bond-angle distortion --- are well established as band-structure engineering knobs (\cite{Chaves2020}). We denote the parent structure at each step as $s$ and the output structure by the LATHE-equipped agent as $s'$. The \emph{design} setting, in contrast, is more relevant for inverse materials design (\cite{SanchezLengeling2018, Zunger2018}): the Hypothesis Agent is required to include energy as an additional minimization goal in every proposal, and the output of the LATHE-equipped agent undergoes additionally an unconstrained relaxation (i.e. both lattices and atomic positions can be altered) with the MLIP at every step to fulfill the requirement of material design that every structure in the design process should sit in an energy basin. We denote the structure after relaxation as $s''$.

For the \emph{manipulation} setting, we accept one step as long as the band gap of $s'$ is strictly closer to the target than that of $s$, while for the \emph{design} setting, we accept one step if all the following three criteria are met: (i) the band gap of $s''$ is strictly closer to the target than that of $s$; (ii) the final structure $s''$ doesn't drift back to the original structure $s$ (as judged by a \texttt{pymatgen StructureMatcher} \cite{Ong2013}); and (iii) the geometric constraints imposed by the LATHE-equipped agent are preserved in a directional way. (i.e. every targeted geometric quantity moved in the desired direction) We didn't adopt a stricter acceptance rule that requires the geometric properties are precisely preserved after relaxation, since an unconstrained relaxation, by construction, almost inevitably moves $s'$ off any purely geometric target imposed by the optimizer: $s'$ is a minimum of the weighted loss, whereas $s''$ is a minimum of solely the energy component. The two coincide only when the geometric target is itself an energy minimum. A precise-preservation rule would therefore reject almost every iteration in which the geometric lever actually did its job. The directional check, despite being a softened approach, still captures what the inverse-design objective requires: that the qualitative intent of the hypothesis (e.g. increase an angle, shorten a bond) survives relaxation.

To expedite exploration in the chemical space, we adopt a stagnation-aware exploration mechanism that, after a configurable number of consecutive non-improving iterations, restarts hypothesis generation from the closest alternate candidate seen so far to escape
unproductive local basins. This is analogous to the perturbation-and-restart strategies used in basin hopping (\cite{basin-hopping}). As a conclusion, these components constitute a hypothesize—optimize—evaluate cycle in which the qualitative materials-science intuition encoded in the Hypothesis Agent is carried out by the LATHE-equipped Agent to yield an interpretable trajectory from a high-level property target to a concrete, dynamically stable optimized structure.

\section{Results}\label{sec:results}

\subsection{\label{sec:single-property-results}Evaluation of LATHE}

\begin{figure}
    \centering
    \includegraphics[width=1\linewidth]{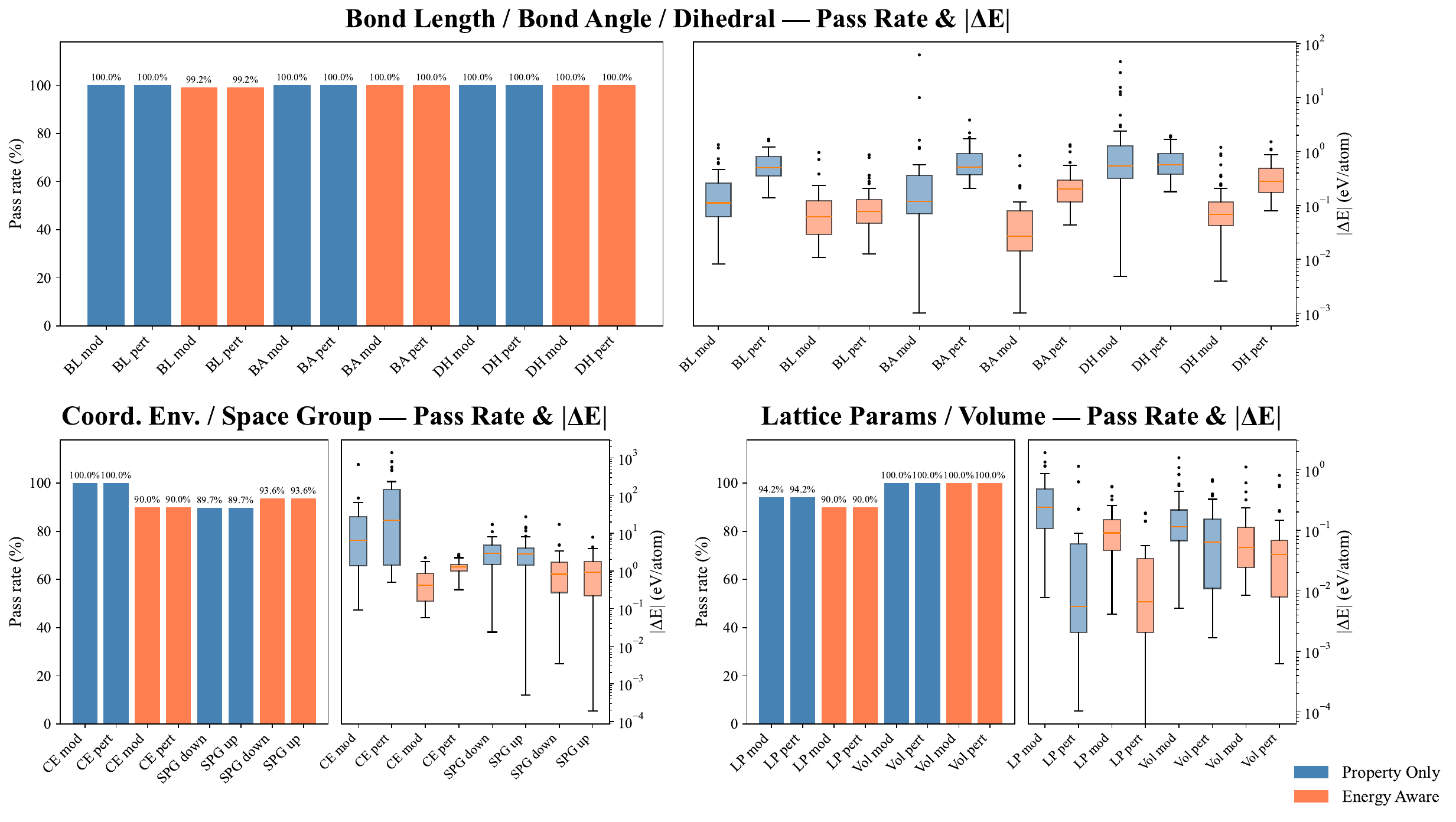}
    \caption{\textbf{Single-property optimization results across seven geometric and symmetry targets.} Pass rates for each property are reported under two settings: property-only optimization (blue) and property jointly optimized with an auxiliary energy loss (orange). LATHE achieves near-perfect pass rates in the property-only setting (100\% for bond length (BL), bond angle (BA), dihedral angle (DH), coordination environment (CE), and volume (Vol); 94.2\% for lattice parameters (LP); 89.7\% for space group (SPG)), and these rates are largely preserved when the energy loss is added, with only modest decreases for coordination environment and lattice parameters. \textbf{Per-atom energy distance $|\Delta E|$ between each optimized structure and its MatterSim-relaxed counterpart, shown on a log scale.} Energy-aware optimization (orange) consistently produces structures closer to a local energy minimum than property-only optimization (blue) across all properties, with the largest reductions observed for coordination environment and space group, confirming that the jointly optimized structures satisfy their geometric targets while lying substantially closer to the potential energy surface minima. Each property is evaluated over 120 test cases (78 for space group) with different test case generating patterns \emph{perturb}/\emph{modify} (\emph{upscale}/\emph{downscale} for space group).}
    \label{fig:single-prop}
\end{figure}

We first evaluate LATHE on seven single-property optimization tasks: bond length, bond angle, dihedral angle, coordination environment, lattice parameters, volume, and space group, with each property tested across 120 cases (78 for space group). For each property except space group, test cases are constructed in two variants: \texttt{perturb}, which applies random noise to the relevant degrees of freedom of a known reference structure and requires the optimization to recover the property in the original structure, and \texttt{modify}, which requires the optimization to achieve properties with  new, randomly sampled target values that differ from the original structure. Space group test cases instead use \texttt{upscale} and \texttt{downscale} variants that raise or lower the target symmetry relative to the input structure. All properties except space group are optimized with direct gradient descent on differentiable loss functions over fractional coordinates and lattice parameters, while space group optimization employs basin-hopping. As shown Figure~\ref{fig:single-prop}, LATHE achieves near-perfect pass rates across all properties in the property-only setting: bond length, bond angle, dihedral angle, coordination environment, and volume each reach 100\%, while lattice parameters and space group attain 94.2\% and 89.7\%, respectively.

When an auxiliary MLIP energy loss (MatterSim, \cite{yang2024mattersim}) is added, pass rates remain largely preserved: bond length, bond angle, dihedral, and volume stay at or above 99\%, and space group has a modest improvement to 93.6\%, with only coordination environment (90.0\%) and lattice parameters (90.0\%) showing moderate decreases, which indicates a mild trade-off between geometric fidelity and energetic awareness for these more constrained properties. To assess whether the energy loss makes the optimization truely energy-aware, we measure the distance from each optimized structure to a local energy minimum by relaxing the optimized structure with MatterSim and computing the per-atom energy difference before and after relaxation ($|\Delta E|$). The energy-aware mode consistently yields lower $|\Delta E|$ across all properties, with particularly dramatic reductions for coordination environment and space group optimizations, which confirms that the jointly optimized structures lie substantially closer to the potential energy surface minima while still satisfying their geometric targets.

\subsection{Performance of the LLM Agents}

\subsubsection{The LATHE-equipped Agent}

\begin{table}[t]
  \centering
  \caption{Representative examples of direct and informative prompts for the same structure (BaTiO$_3$,
  mp-5020).}
  \label{tab:prompt_examples}
  \small
  \begin{tabularx}{\textwidth}{cX}
  \toprule
  \textbf{Style} & \textbf{Prompt} \\
  \midrule
  Direct & Set the bond length between Ti at site 1 and O that is closest to it to 1.95~\AA. \\ & Set the dihedral
  angle of O--Ti--O--Ba where Ti is at site 1, O that is closest to Ti is at site 2, and O that is furthest from
  Ti is at site 4 to 180~deg. \\ & Set the bond angle of O--Ti--O where Ti is at site 1, O is at site 2 and O is
  closest to Ti to 180.0~deg. Minimize the MLIP-predicted energy of the structure. \\
  \midrule
  Informative & Tuning the local oxygen coordination around the A- and B-site cations in perovskite BaTiO$_3$
  offers a pathway to control lattice polarization and ferroelectric instability, thereby shifting
  phase-transition temperatures and enhancing dielectric and piezoelectric performance for actuators and sensors. Increase some Ba--Ti--O bond angles in the structure by at least 10 degrees. Increase some Ti--O bond lengths
  in the structure by at most 5\%. \\
  \bottomrule
  \end{tabularx}
\end{table}

\paragraph{Test Case Creation} To evaluate the agent's ability to translate natural-language structure-modification requests into executable optimization configurations, we construct a benchmark of 40 prompts with a template-driven pipeline over 10 input structures. For each prompt, a random compatible subset of the seven supported properties is sampled, an MLIP energy objective is appended with 30\% probability, and an LLM (\texttt{GPT-5.4-nano}) fills the templates with chemically plausible values conditioned on the formula and CIF (SI~\ref{si:benchmark}). Each case is generated in two styles. \textit{Direct} prompts specify targets unambiguously through precise numerical values, site indices, and property names (e.g. ``Set the bond length between Ti at site 1 and O that is closest to it to 1.95~\AA.''). \textit{Informative} prompts introduce deliberate ambiguity: relative modifications with qualitative quantifiers (e.g. ``Increase some Ti--O bond lengths by at most 5\%''), prefixed by a sentence of scientific context that motivates the modification. This style reflects how the LATHE-equipped agent would receive hypotheses or instructions from other agents in a multi-agent crystal design setting. Table~\ref{tab:prompt_examples} gives representative examples of each style.

\paragraph{LLM-as-Judge Evaluation} We judge success with a three-stage pipeline. An LLM \emph{parser} converts the prompt and the \emph{original} structure into a typed checklist of requirements, each carrying a property kind, site indices, a target value, a relative-change encoding, and a quantifier. A \emph{deterministic verifier} (pure Python, no LLM) then evaluates every requirement with MIC geometry helpers, tolerance-aware comparators, and element-tuple enumeration for non-specific quantifiers, re-resolving relative descriptors against the original structure and overriding the parser on disagreement; it returns PASS, FAIL, SKIP, or INCONCLUSIVE and is authoritative whenever conclusive. Only if a non-energy requirement remains inconclusive do we invoke an \emph{LLM verifier} backup, which receives the prompt, the checklist, and both CIFs, and resolves the remaining requirements using a sandboxed \texttt{run\_python} tool. Full prompts, the requirement schema, and tolerance settings are given in SI~\ref{si:judge}.

\begin{table}[h]
\centering
\caption{Performance comparison of the three methods on the 40-task benchmark.}
\label{tab:baseline_comparison}
\begin{tabular}{lcccc}
\toprule
\textbf{Method} & \textbf{Direct CIF Generation} & \textbf{Claude Code} & \textbf{Ours} \\
\midrule
\textbf{Success Rate} & 45.0\% & 72.5\% & \textbf{87.5\% $\pm$ 3.39\%} \\
\bottomrule
\end{tabular}
\end{table}

\paragraph{Experimental Outcomes} We similarly used \texttt{gpt-5.4-nano} as the backbone LLM for the agent. Across 200 runs (40 prompts $\times$ 5 independent trials with different seeds per prompt), the agent achieves an overall success rate of 87.5\%. Performance is related to prompt phrasing: direct prompts pass at 93.0\%, while informative prompts, despite their deliberate ambiguity and qualitative quantifiers, pass at 82.0\%, which indicates that the agent can ground relative and underspecified language into concrete numerical targets. Of the 25 failures, 24 (96\%) are attributable to optimization failure, in which the agent correctly interprets the prompt and assembles a valid configuration but the underlying gradient-based optimizer cannot simultaneously satisfy all constraints; only 1 failure (4\%) stems from intent misunderstanding, where the agent misidentifies the user's geometric target or omits a required constraint. This suggests that the LLM-driven translation layer is reliable in understanding the structure and compiling the abstract requirements into concrete specifications for the optimization tools. Yet, it sometimes ignores whether the provided optimization targets are feasible so that the subsequent optimization may fail.

\paragraph{Baseline Methods} To establish a baseline, we evaluated the performance of two alternative approaches. \textbf{(1) Direct CIF generation:} we instruct the LLM (\texttt{gpt-4o}) to directly generate the CIF representation of the transformed structure given the input CIF and the task prompt. \textbf{(2) Claude Code:} we provide the task prompt, the path to the input structure, and a Python environment with \texttt{ase}, \texttt{pymatgen}, and other scientific computing libraries preloaded, then instruct Claude Code (\texttt{claude-opus-4-6}) to execute the LATHE-equipped by following an observe-modify-confirm workflow. These two methods represent the naive LLM solution and the state-of-the-art agentic coding approach, respectively. We leveraged the same judgment workflow as introduced above. Direct CIF generation succeeds on 18/40 (45\%) tasks, and Claude Code succeeds on 29/40 (72.5\%) tasks, both inferior to our LATHE-equipped agent. Moreover, while Claude Code achieves reasonable performance, it exhibits two key limitations relative to our approach. First, Claude Code must recompose tailored scripts for every step of the pipeline, both for observing the structure and for executing the geometric operation, whereas our method, despite also requiring code execution at the observation stage, incurs no additional scripting overhead during optimization, as LATHE handles it natively. Second, Claude Code struggles to reconcile geometric constraints with the energy co-objective. In one failure mode, it first satisfies the geometric constraints and then performs a relaxation to reach the energy minimum, which frequently breaks the previously satisfied geometry; Claude Code must then re-enforce the geometric constraints, and this back-and-forth cycle can repeat for several iterations, resulting in substantial time and token cost. In a second failure mode, Claude Code attempts to incorporate energy directly into the constraint satisfaction step; however, since it seeks an energy minimum, it repeatedly rewrites the optimization script with different implementations and hyperparameters, again producing a multi-turn and token-intensive process. We observe that these two failure modes, individually or in combination, can drive individual Claude Code runs to exceed 30 minutes with unacceptably high token costs.

\subsubsection{A Case Study: Structural Design for Optimizing Band Gap}

\begin{figure*}[t]
\centering
\includegraphics[width=\linewidth]{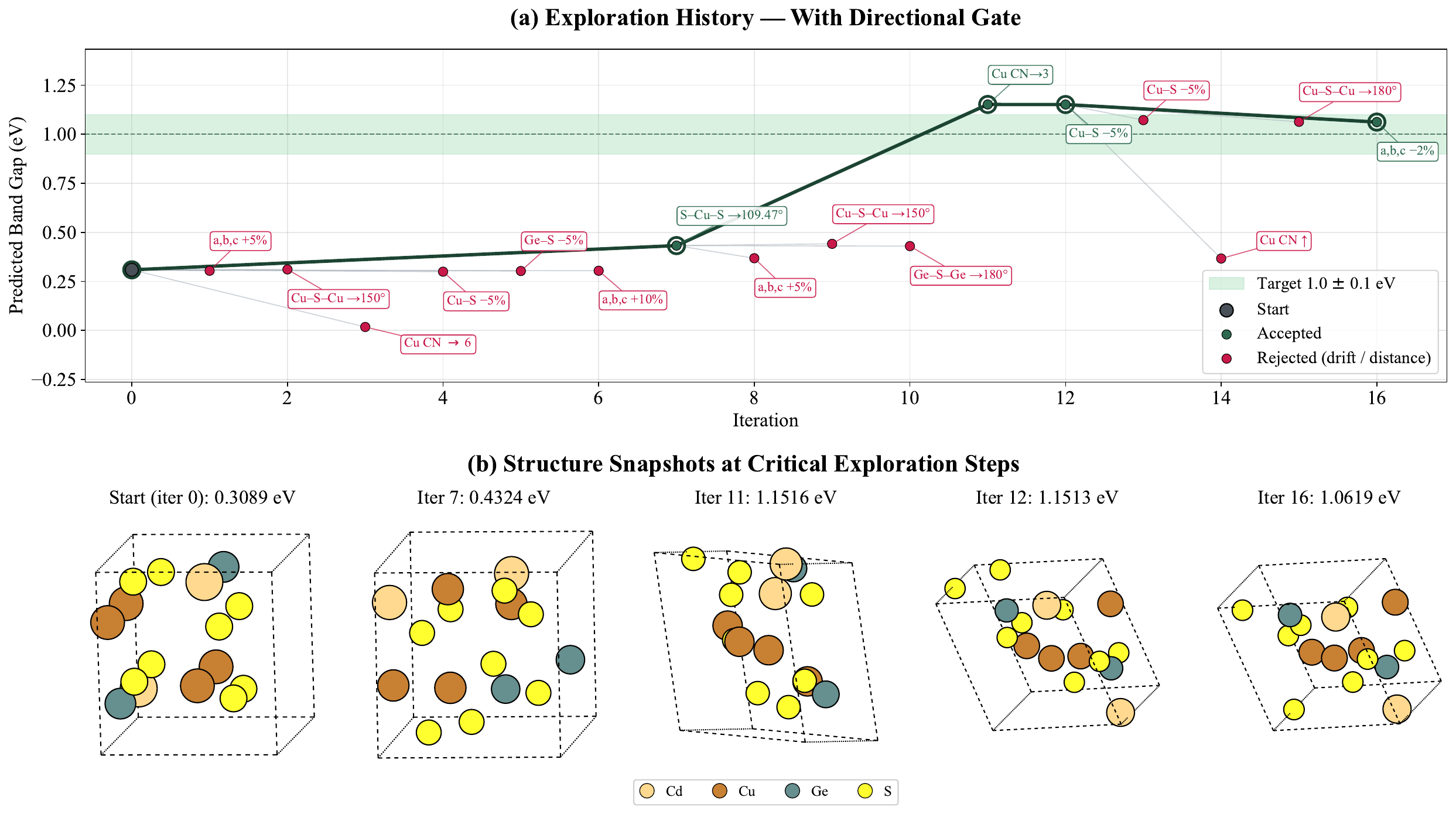}
\caption{\textbf{Example \texttt{design} run of the multi-agent design loop on Cu\textsubscript{2}CdGeS\textsubscript{4} (mp-13982), targeting a band
gap of $1.0 \pm 0.1$~eV.} (a) Branching search tree of the run. Each node is one iteration, plotted at $(i, \,g_i)$ where $g_i$ is the post-relaxation MatterSim-predicted band gap returned by the relaxation gate; the starting structure sits at $(0,\,0.309~\mathrm{eV})$. Edges connect each iteration to the parent from which it was branched off. Green markers denote candidates accepted by the acceptance rule, and the red markers denote rejections. The thick green path marks the lineage of accepted ancestors that produced the final structure. Each node is annotated with a short tag for the geometric lever the Hypothesis Agent proposed. The dashed horizontal line marks the target band gap and the shaded band the $\pm 0.1$~eV tolerance window. (b) Crystal-structure snapshots along the accepted backbone: the input structure (iter~0), the iteration~7 structure after the S--Cu--S angles were pushed toward the tetrahedral ideal of $109.47^{\circ}$, the iteration~11 structure after the Cu coordination was lowered from~$4$ to~$3$, the iteration~12 structure after a $-5\%$ contraction of the Cu--S bonds, and the final iteration~16 structure after a $-2\%$ isotropic lattice contraction that lands the band gap inside the target window.}
\label{fig:band-gap-results}
\end{figure*}

\begin{table*}[t]
\centering
\small
\setlength{\tabcolsep}{4pt}
\begin{tabular}{c c p{0.30\linewidth} c c c c}
\toprule
Iter & Parent & Hypothesis (summarized) & Raw gap (eV) & Relaxed gap (eV) & Decision \\
\midrule
1& Start& Increase $a, b, c$ by $\sim\!5\%$. & 0.287 & 0.304 & rejected (drift) \\
2& Start& Tilt Cu--S--Cu bond angles toward $150^{\circ}$. & 0.417 & 0.311 & rejected (drift) \\
3& Start& Raise Cu coordination from~$4$ to~$6$. & 0.051 & 0.018 & rejected (distance) \\
4& Start& Shorten Cu--S bond lengths by $\sim\!5\%$. & 0.267 & 0.300 & rejected (drift) \\
5& Start& Lengthen Ge--S bond lengths by $\sim\!5\%$. & 0.280 & 0.303 & rejected (drift) \\
6& Start& Increase $a, b, c$ by $\sim\!10\%$. & 0.804 & 0.304  & rejected (drift) \\
7& Start& Open S--Cu--S bond angles toward $109.47^{\circ}$. & 0.444 & \textbf{0.432} & \textbf{accepted} \\
8& Iter 7 & Increase $a, b, c$ by $\sim\!5\%$. & 0.394 & 0.368 & rejected (distance) \\
9& Iter 7 & Tilt Cu--S--Cu bond angles toward $150^{\circ}$. & 0.472 & 0.441 & rejected (drift) \\
10 & Iter 7 & Open Ge--S--Ge bond angles toward $180^{\circ}$. & 0.430 & 0.430 & rejected (drift) \\
11 & Iter 7 & Lower Cu coordination from~$4$ to~$3$. & 0.841 & \textbf{1.152} & \textbf{accepted} \\
12 & Iter 11 & Shorten Cu--S bond lengths by $\sim\!5\%$. & 1.088 & \textbf{1.151} & \textbf{accepted} \\
13 & Iter 12 & Shorten Cu--S bond lengths by $\sim\!5\%$. & 1.159 & 1.072 & rejected (drift) \\
14 & Iter 12 & Raise Cu coordination. & 0.117 & 0.367 & rejected (distance) \\
15 & Iter 12 & Open Cu--S--Cu bond angles toward $180^{\circ}$. & 1.170 & 1.063 & rejected (drift) \\
16 & Iter 12 & Reduce $a, b, c$ by $\sim\!2\%$. & 1.065 & \textbf{1.062} & \textbf{accepted (target)} \\
\bottomrule
\end{tabular}
\caption{\textbf{Per-iteration record of the example \texttt{design} run on Cu\textsubscript{2}CdGeS\textsubscript{4} (mp-13982).} Each row reports the parent structure the Hypothesis Agent branched from (``Start'' denotes the input structure), the natural-language hypothesis it proposed (the trailing ``minimize the MLIP-predicted energy'' clause that appears in every proposal under the directional gate is omitted from the table), the band gap of the gradient-optimized geometry (raw), the band gap after a brief MLIP relaxation through the relaxation gate (relaxed), and the directional-rule decision. Rejections are labeled ``drift'' (the relaxed structure $s''$ matches the iteration input $s$), ``distance'' ($s''$ differs from $s$ but the gap moved further from the
target), or ``directional'' (the geometric property enforced during optimization doesn't hold directionally after the relaxation). The starting band gap is $0.309$~eV and the target is $1.0 \pm 0.1$~eV. The run terminates after iteration~16, when the post-relaxation gap of $1.062$~eV enters the target window.}
\label{tab:bandgap_run02}
\end{table*}

\paragraph{Walking through an example run.} To illustrate how the two-agent system with the hypothesize--translate--optimize--relax--evaluate loop introduced in Section~\ref{sec:multi-agent-design} works in practice, we walk through a single representative \texttt{design} run targeting a band gap of $1.0 \pm 0.1$~eV on Cu\textsubscript{2}CdGeS\textsubscript{4} (with MPID \texttt{mp-13982}), as the \texttt{design} mode resembles structure design better. The starting structure has a MatterSim-predicted band gap of $0.309$~eV, so the agent must roughly triple the gap purely through geometric modifications proposed in natural language and executed by the downstream optimization back-end. Acceptance is governed by the \emph{directional} gate. We denote the original structure at each iteration as $s$, the structure right after optimization as $s'$, and the post-optimized structure after an unconstrained geometric relaxation as $s''$. A candidate is admitted only if its post-relaxation structure $s''$ differs from the iteration input $s$, moves the band gap closer to the target, and each targeted geometric quantity (together with the band gap itself) shifts away from the source-relaxed reference in the direction the Hypothesis Agent explicitly predicted. The Hypothesis Agent selects each iteration's parent from a ranked candidate pool, so adjacent iterations are not necessarily causally linked; the full exploration unfolds as a branching tree (Figure~\ref{fig:band-gap-results}) rather than a linear sequence, and Table~\ref{tab:bandgap_run02} records the parent--hypothesis--verdict for each iteration. We use \texttt{gpt-5.4-nano} as the backbone LLM for the LATHE-equipped Agent and \texttt{gpt-4o} as the backbone LLM for the Hypothesis Agent.

Iterations~1--6 all branch off the input structure and are all rejected, exposing two distinct failure modes that the directional gate exists to catch. Iterations~1, 2, 4, 5, and~6 propose isotropic lattice expansion ($+5\%$, then $+10\%$), Cu--S--Cu angle tilting toward $150^{\circ}$, and Cu--S / Ge--S bond-length perturbations; in each case the gradient back-end returns an $s'$ whose raw band gap is sometimes large, but the subsequent relaxation  collapses the structure straight back to the input, so the ``drifted-back'' rule rejects them outright (red markers at iters~1, 2, 4, 5, 6 in Figure~\ref{fig:band-gap-results}a). After six straight rejections, the escalation block of the Hypothesis Agent prompt actively steers the proposer away from attempts that have already failed.

Iteration~7 is the first acceptance: opening the S--Cu--S bond angles toward the tetrahedral ideal of $109.47^{\circ}$ produces a structure that survives relaxation as a genuinely different geometry, moves the gap to $0.432$~eV, and passes the directional gate (the targeted bond-angle quantity moves in the predicted direction to $109.47^{\circ}$. Now iterations~8--11 branch from iteration~7 rather than from the input. Iterations~8 ($+5\%$ lattice), 9 (Cu--S--Cu tilt to $150^{\circ}$), and~10 (Ge--S--Ge tilt to $180^{\circ}$) are rejected---the first because the relaxed gap of $0.368$~eV moves further from target than the parent's $0.432$~eV, the latter two because they drift back to the iteration~7 basin under MLIP relaxation. Iteration~11 then proposes a qualitatively new lever: \emph{lower} the Cu coordination from~$4$ to~$3$. This move lands in a distinct basin (no RMS match) and the post-relaxation gap jumps to $1.152$~eV, overshooting the target window by a small amount but passing the directional gate. Iteration~11 is accepted and becomes the new working candidate. The final two acceptances refine this candidate downward toward the target window. From iteration~11, iteration~12 contracts the Cu--S bonds by $\sim\!5\%$; the relaxed structure remains distinct ($\mathrm{RMS}\approx 0.080$~\AA), the gap holds at $1.151$~eV, and iteration~16 finally proposes a small $-2\%$ isotropic lattice contraction; the relaxed structure differs from the parent ($\mathrm{RMS}\approx 0.064$~\AA), the lattice contracts as predicted in direction, and the post-relaxation gap of $1.062$~eV lands inside the $1.0 \pm 0.1$~eV target window. The run terminates after sixteen iterations along a five-node accepted backbone (Start $\to$ 7 $\to$ 11 $\to$ 12 $\to$ 16; bold green path in Figure~\ref{fig:band-gap-results}a; corresponding snapshots in Figure~\ref{fig:band-gap-results}b).

\paragraph{Robustness across random seeds.} The trajectory analyzed above is one realization of a stochastic loop: the Hypothesis Agent samples from a language-model policy, and the downstream gradient optimizer has its own randomness. To test whether the multi-agent system is robust to this combined noise, we repeat the same Cu\textsubscript{2}CdGeS\textsubscript{4} band-gap task ($1.0 \pm 0.1$~eV target on \texttt{mp-13982}) for ten independent runs in each of the two settings, under identical settings and a budget of 50 iterations per run. Both settings converge in nine of the ten runs. Among the successful runs, the \emph{design} setting reaches the tolerance window in a median of $13$ iterations (mean $14.9$), while the \emph{manipulation} setting requires a median of $16$ iterations (mean $19.0$). The single failed run of the design setting exhausts its budget at $1.16$~eV — overshooting the upper edge of the tolerance window, and the failed run of the manipulation setting stalls at $0.87$~eV, still moving in the right direction but unable to close the remaining $0.13$~eV gap within the iteration budget. We attribute the design setting's tighter iteration distribution to the relaxation gate and the directional check: by accepting only steps whose qualitative intent survives an MLIP relaxation, the loop suppresses brittle changes that immediately drift back and guides the search with informative signals from an informative exploration history, in which each step fulfills the hypothesis proposed by the hypothesis agent. The \texttt{manipulation} setting, in contrast, accepts any geometric configuration that lowers the unrelaxed band gap, which makes the per-iteration acceptance rate higher but the trajectory noisier: several runs spend much of their budget exploring high-strain configurations that a relaxation
would have undone. In conclusion, these statistics indicate that the closed loop reliably solves the band gap optimization task under randomness in both settings, and that gating on MLIP-relaxed, directionally-consistent candidates, adhering to principles of inverse design. does not degrade convergence efficiency.

\subsection{Discussion}

As a summary, the single-property experiments establish that expressing crystallographic properties as differentiable losses over the fractional coordinates and lattice matrix is sufficient to drive gradient-based optimization to near-perfect constraint satisfaction across a heterogeneous set of geometric targets. The uniformly high pass rates for single-property optimizations indicate that the loss formulations produce well-behaved gradient landscapes that standard first-order optimizers can navigate without additional tuning. Furthermore, optimization with energy awareness preserves these pass rates while substantially reducing the distance from the optimized structure to a local energy minimum, which demonstrates that the auxiliary energy loss functions successfully a regularizer that tightens physical plausibility without overriding the geometric objective. The LATHE-equipped Agent inherits this reliability: across 200 runs it translates natural-language prompts into executable configurations with a 87.5\% success rate, a performance surpassing the SOTA coding agent Claude Code. This shows that the agent can ground qualitative quantifiers such as ``some'' or ``at most 5\%'' into concrete numerical targets. That most observed failures arise from optimization-side infeasibility rather than translation-side misunderstanding is the most informative statistic in this evaluation: the language-to-configuration layer is not the bottleneck, and further gains will come from the optimizer and from better feasibility reasoning, not from a larger or more capable LLM translator.

The band-gap case study demonstrates that the hypothesize–translate–optimize–evaluate loop is capable of solving a nontrivial inverse-design task. Nine of ten independent \texttt{design} runs on Cu$_2$CdGeS$_4$ reach the $1.0 \pm 0.1$ eV target within the 50-iteration budget. This behavior emerges from in-context conditioning on the running history of (hypothesis, modification, outcome) triples, without any task-specific fine-tuning. Equally important, the division of labor insulates each agent from the failure modes of the other: the Hypothesis Agent need not reason about gradient landscapes or optimizer hyperparameters, and the LATHE-equipped Agent need not reason about band-gap physics. This design yields a fully auditable trajectory in which every accepted structural change is associated with a human-readable hypothesis and a measured outcome.

Two directions stand out for extending the present framework. First, the LATHE-equipped Agent currently converts each requested constraint to a loss covered by a validator, but does not perform a feasibility analysis across the full set of targets on a given structure. This gap is most visible for ambiguous prompts, where qualitative quantifiers such as ``some'' or ``at most'' are grounded into concrete numbers without an accompanying feasibility check, which sometimes results in an over-demanding set of constraints. Future work could add a feasibility pre-screening step, for instance using cheap geometric or symmetry-based heuristics to flag jointly unsatisfiable constraint sets before optimization begins, or to suggest relaxed targets when a strict combination cannot be met. Second, because the gradient-based optimizer is intrinsically local, jointly satisfying a non-trivial set of geometric constraints and reaching a low-energy MLIP basin can lead to a compromise between the two objectives, particularly when their minima lie in different regions of configuration space. A promising avenue here is to pair the framework with generative crystal models that sample directly from stable-structure distributions, conditioning them on differentiable constraint terms during sampling or hybridizing them with a local optimizer that enforces geometric targets as a post-hoc projection step. Realizing this would let future systems reach precisely-constrained, energy-favorable structures beyond what local optimization alone can access, and would extend the hypothesize–optimize–relax loop to targets that are currently out of reach.

\section{Conclusion}\label{sec:conclusion}

We introduced LATHE, a geometric reasoning toolkit for hypothesis-driven crystal structure exploration that bridges the gap between qualitative chemical intuition and quantitative geometric control within a single, language-accessible, gradient-powered framework. At the algorithm level, seven classes of crystallographic properties --- bond length, bond angle, dihedral angle, coordination environment, lattice parameters, cell volume, and space group --- are expressed as differentiable losses over the fractional coordinates and lattice matrix so that any arbitrary combinations of targets can be optimized via direct gradient-based optimization. Exposing LATHE through an MCP server turns it into a service that autonomous agents can discover, compose against, and invoke, and the LATHE-equipped Agent built on this interface translates natural-language structure-modification requests into executable optimization configurations with high fidelity across both direct and deliberately ambiguous prompts. Embedding this agent inside a closed-loop multi-agent system together with a Hypothesis Agent further lifts the framework from structural editing to property-oriented inverse design to produce an interpretable, auditable trajectories in which every accepted structural change is associated with a human-readable hypothesis and a measured property outcome. Together, these results show the feasibility of a promising  mode of computational materials design in which fine-grained geometric operations become the primitive building blocks of an LLM-driven discovery loop, opening a path toward systems that reason about structure, execute on it with physical fidelity, and explain their trajectory in the same language in which the design problem was first posed.

\section{Data Availability}
The authors declare that all data, materials, and code supporting the results reported in this study are available at the following repository: \url{https://github.com/Fung-Lab/StructGradDesign}.

\section{Acknowledgements}
We gratefully acknowledge the support of a Google Research Scholar Award.




\clearpage
\bibliography{refs}
\bibliographystyle{plainnat}

\clearpage

\appendix
\counterwithin{figure}{section}
\counterwithin{table}{section}
\counterwithin*{equation}{section}
\renewcommand\theequation{\thesection\arabic{equation}}

\begin{titlepage}
  \centering
  \LARGE \textsc{Supplementary Information} \par
  \let\endtitlepage\relax
\end{titlepage}

\section{Implementation Details for the Geometric Loss Functions}\label{sec:SI-loss-details}

\subsection{Coordination Environment}\label{sec:SI-coord-env}

For each neighboring atom $j$ at distance $r$ from the target site, its contribution to the soft coordination count is computed as
$$c_j = \frac{1}{2}\left[\text{erf}\!\left(\frac{r_\text{cut} - r}{\sigma\sqrt{2}}\right) - \text{erf}\!\left(\frac{r_\text{min} - r}{\sigma\sqrt{2}}\right)\right] \cdot \sigma\!\left(\frac{r_\text{cut} - r}{w}\right)$$
where $\sigma$ is the Gaussian smearing width, $r_\text{min}$ excludes unphysical self-contacts, and the trailing sigmoid term with envelope width $w$ acts as a Fermi-like taper that smoothly suppresses contributions near the cutoff boundary. The soft coordination number for a given species is $\tilde{n} = \sum_j c_j$.
 
Three auxiliary penalty terms are added to the primary $(\tilde n - n^*)^2$ loss:
 
\begin{enumerate}
\item \textbf{Sharpness penalty}: $\sum_j c_j(1 - c_j)$, maximized when contributions are near $0.5$ (ambiguous membership) and near zero when each atom is clearly inside or outside the coordination shell. This term is gated by a Gaussian proximity weight $\exp(-2(\tilde{n} - n^*)^2)$ so that it only acts when the count is already near the target.
\item \textbf{Integer-snapping penalty}: $\frac{1}{2}(1 - \cos(2\pi\tilde{n}))$, a smooth periodic potential with non-zero gradient at all fractional counts, eliminating gradient dead zones near integer boundaries. This term is likewise gated by the proximity weight.
\item \textbf{Soft-core repulsion and attractive well}: a soft-core repulsion $\sum_j \max(r_\text{rep} - r_j,\, 0)^2$ penalizes any neighbor closer than a minimum contact distance $r_\text{rep}$, preventing unphysically collapsed configurations. To ensure that under-coordinated sites can recover when atoms drift beyond the hard cutoff, an attractive well potential acts on atoms within an extended shell $(r_\text{cut},\, r_\text{cut} + \Delta r]$, contributing a quadratic restoring force scaled by the coordination deficit $\max(n^* - \tilde{n}, 0)$.
\end{enumerate}

\subsection{Wyckoff Site Warm-Start Initialization}\label{sec:SI-wyckoff-initialization}

Given the target space group and the atomic composition of the structure, the initializer first determines all valid decompositions of each species' atom count into sums of Wyckoff multiplicities, then searches over these decompositions to find the placement that minimizes a weighted combination of atomic displacement from the original coordinates and symmetry compliance cost. For each candidate decomposition, generator points for parametric Wyckoff sites (those with one or more degrees of freedom) are optimized via L-BFGS-B (\cite{Zhu1997-cq}) over the free fractional coordinates, with Hungarian matching used to assign original atomic positions to proposed orbit positions in a displacement-minimizing way. Fixed special sites (zero degrees of freedom) are placed directly at their crystallographic positions. Placements that produce interatomic distances below a physical minimum threshold, or that overlap with already-placed species, are discarded. When the primitive cell atom count is incompatible with any Wyckoff decomposition of the target space group, the cell is expanded into a supercell by the smallest diagonal scaling matrix whose atom counts admit a valid decomposition, and constraint site indices, if additional constraints that require atomic indices exist, are remapped accordingly. Across all valid decompositions, the final placement is selected by Hungarian matching as the one that minimizes total atomic displacement from the original structure. This ensures that the warm-started configuration is as close as possible to the input geometry while satisfying the Wyckoff site constraints of the target space group. This displacement-minimizing initialization distinguishes our approach from generative methods that synthesize a symmetry-compliant structure from scratch (\cite{jiao2024crystalstructurepredictionjoint, kelvinius2025wyckoffdiffgenerativediffusion, levy2025symmcdsymmetrypreservingcrystalgeneration,  Fredericks_2021}): rather than producing an arbitrary representative of the target space group, our initializer anchors the symmetry-constrained configuration to the user's input structure so that the geometry of the original structure is preserved as much as possible before gradient-based optimization begins.

\section{Benchmark Experiment Designs}

\subsection{Natural Language-Based Geometric Optimization Test Case Creation} \label{si:benchmark}

\begin{table}[t]
\centering
\small
\caption{Sentence skeletons per property. Bracketed items are filled by the LLM;
braced items are drawn by the pipeline. \emph{Informative} prompts use the
ambiguous skeletons.}
\label{tab:si_templates}
\begin{tabular}{@{}lp{4.6cm}p{4.6cm}@{}}
\toprule
Property & Direct & Ambiguous \\
\midrule
Bond length & Set the bond length between [X] \{q$_1$\} and [Y] \{q$_2$\} to [value]. & \{op\} \{quant\} [X]--[Y] bond lengths by \{magnitude\}. \\
Bond angle & Set the bond angle of [X]--[Y]--[Z] where [Y] is at site \{$i$\}, [X] \{q$_x$\}, and [Z] \{q$_z$\} to [value]. & \{op\} \{quant\} [X]--[Y]--[Z] bond angle(s) by \{magnitude\} degrees. \\
Dihedral & Set the dihedral angle of [X]--[Y]--[Z]--[W] where [Y] is at site \{$i_y$\}, [Z] is at site \{$i_z$\}, [X] \{q$_x$\}, and [W] \{q$_w$\} to [value]. & \{op\} \{quant\} [X]--[Y]--[Z]--[W] dihedral angle(s) by \{magnitude\} degrees. \\
Coordination & Set the coordination number of [X] at site \{$i$\} to [$n$]; \emph{or} set the number of [X] atoms near [Y] at site \{$i$\} to [$n$] within \{$d$\}~\AA. & \{quant\} [X] atom(s) should be [$n$]-coordinated; \emph{or} \{op\} the number of [X] atoms near \{quant\} [Y] atom within \{$d$\}~\AA. \\
Lattice & Set lattice parameter(s) [$a$$\vert$$b$$\vert$$c$$\vert$$\alpha$$\vert$$\beta$$\vert$$\gamma$] to [value(s)]. & \{op\} \{quant\} lattice length(s)/angle(s) by \{magnitude\}. \\
Volume & Set the volume of the unit cell to [value]. & \{op\} the volume of the unit cell by \{magnitude\}. \\
Symmetry & Set the space group to [symbol / ITA number]. & Change the space group to one with higher (lower) symmetry, or to one in the [crystal system] system. \\
Energy & \multicolumn{2}{l}{Minimize the MLIP-predicted energy of the structure. (fixed sentence)} \\
\bottomrule
\end{tabular}
\end{table}

Each benchmark prompt is produced by a four-stage pipeline: (i) sample a set of target properties, (ii) instantiate a parameterized sentence template per property, deterministically fixing all atom-selection arguments, (iii) call an LLM to fill only the chemical and numerical blanks, and (iv) validate the returned sentence and re-roll on failure. Only stage (iii) involves an LLM; the site indices, anchors, and distance cutoffs that determine \emph{which} atoms a requirement refers to are drawn by the pipeline itself, so every prompt has a well-defined referent independent of model behavior.

\paragraph{Property sampling} The structural property pool is \{\texttt{bond\_length}, \texttt{bond\_angle}, \texttt{dihedral\_angle}, \texttt{coordination}, \texttt{volume}, \texttt{symmetry}, \texttt{lattice\_params}\}. For each prompt we draw two properties without replacement, with sampling weights $1.0$ for all properties except \texttt{symmetry} ($0.25$) and \texttt{coordination} ($0.75$); these two are down-weighted because they act as global constraints that dominate the optimization. Three compatibility rules are then applied in order: if \texttt{symmetry} was drawn, it becomes the sole structural target. Otherwise, if \texttt{coordination} was drawn, it becomes the sole structural target. If both \texttt{lattice\_params} and \texttt{volume} were drawn, \texttt{volume} is dropped, since the two are redundant. Independently, an MLIP energy objective (``Minimize the MLIP-predicted energy of the structure.'') is appended with probability of 0.5. A prompt therefore contains one or two structural requirements plus an optional energy requirement.

\paragraph{Template instantiation}
Table~\ref{tab:si_templates} lists the skeleton used for each property and style. For the three properties whose direct form requires atom selection (\texttt{bond\_length}, \texttt{bond\_angle}, \texttt{dihedral\_angle}), the selector slots are filled by the pipeline before the LLM is called, drawing uniformly from five selector families: \emph{explicit index} (i.e. ``is at site $i$''), \emph{nearest / farthest} (i.e. ``is the atom closest (furthest) to (from) the atom at site $a$''), \emph{distance-bounded} (i.e. ``is the atom whose current distance to the atom at site $a$ is smaller (larger) than $d$~\AA'').

\paragraph{LLM blank-filling.}
The instantiated template, the chemical formula, and the full CIF of the starting structure are passed to \texttt{<MODEL>} at temperature $T=0.9$. The system prompt restricts the model to emitting a single completed sentence and forbids copying template scaffolding (placeholder tokens, slash alternations such as ``closest/furthest'', bare atom letters \texttt{X}/\texttt{Y}/\% \texttt{Z}/\texttt{W}, or meta-language). The model's only degrees of freedom are the element symbols and the numerical targets, which it is instructed to choose so as to be chemically plausible for the given formula. One sentence is generated per property and the sentences are concatenated to form the prompt.

\paragraph{Styles.}
\textit{Direct} prompts use the direct skeletons: unambiguous targets specified by exact site indices, property names, and precise numerical values. \textit{Informative} prompts use the ambiguous skeletons, in which the target is expressed as a relative modification (\{increase, reduce\} by \{at most, at least\} $x\%$ or $x$ degrees) over a set of atoms picked out by a qualitative quantifier (\{one, two, three, some, all\}). A second LLM call then produces one sentence of scientific motivation for the sampled modification, which is prepended to the prompt; that call is explicitly instructed not to restate or paraphrase the modification, so the context sentence adds framing without disambiguating the target.

\paragraph{Validation and re-rolling.}
Each generated sentence is checked by a deterministic validator that rejects: (i) residual placeholder tokens; (ii) unresolved slash alternations; (iii) echoed template meta-language; (iv) bare \texttt{X}/\texttt{Y}/\texttt{Z}/\% \texttt{W} atom letters; (v) grammar leaks in which the template's copula is retained in front of a selector that carries its own verb (e.g. ``Ca is whose current distance\ldots''); (vi) repeated declared site indices or an atom pinned by both an index and a relational qualifier; and (vii) for bond angles, failure to enumerate three separate atom selectors, which would indicate that two qualifiers had been fused onto a single outer atom. On rejection the template is re-instantiated with freshly drawn selectors and the call is repeated, for up to four attempts in total; the failure reason and the offending sentence are logged. Prompts that exhaust all attempts are recorded and discarded.

\subsection{LLM-as-Judge Evaluation Protocol}
\label{si:judge}

Every run is scored by the same three-stage pipeline, which consumes the prompt,
the original CIF, and the modified CIF and returns a per-requirement checklist
plus a binary judgement. Only Stages~1 and~3 use an LLM.

\begin{table}[t]
\centering\small
\caption{Tolerances used by the deterministic comparators.}
\label{tab:si_tolerances}
\begin{tabular}{@{}lc@{}}
\toprule
Property & Tolerance \\
\midrule
Bond length, lattice length & $0.1$~\AA \\
Bond angle, lattice angle & $1.0^\circ$ \\
Dihedral angle & $2.0^\circ$ \\
Volume & $5.0$~\AA$^3$ \\
Space group, coordination & exact match \\
\bottomrule
\end{tabular}
\end{table}

\paragraph{Stage 1 --- Parser}
A single structured-output LLM call converts the prompt into an ordered list of typed \texttt{Requirement} objects. Each carries a property kind (\{bond\_length, bond\_angle, dihedral\_angle, coordination, lattice\_length, lattice\_angle, volume, space\_group, energy\}), site indices into the \emph{original} structure, a target value, a relative-change encoding (absolute, exact delta, or an at-most/at-least bound in a given direction, with an optional percent flag), and a quantifier (\{specific, some, all, count\}). The parser sees only a compact summary of the original structure --- lattice parameters, volume, and one line per site giving index, element, and fractional
coordinates.

The convention is that every relative phrase in a prompt (``closest to'', ``furthest from'', ``within $D$~\AA'') denotes an atom in the \emph{original} structure, since the atom nearest a given anchor before modification need not be the nearest one after. The parser therefore resolves such phrases immediately into integer indices, and the downstream verifier never re-interprets prompt wording. Ties are broken by lowest index, with all tied alternatives recorded so that moving any of them counts as satisfying the requirement.

\paragraph{Stage 2 --- Deterministic verification}
Each requirement is re-checked in pure Python and assigned PASS, FAIL, SKIP, or INCONCLUSIVE. All geometry uses minimum imageconvention, and comparisons are tolerance-aware (Table~\ref{tab:si_tolerances}).

For non-specific quantifiers, all element-matching tuples are enumerated under a neighbour cutoff retried at $4.0 \to 5.0 \to 6.0$~\AA. Writing $T$ for the enumerated tuples and $H \subseteq T$ for those satisfying the bound, \texttt{some} passes iff $|H| \geq 1$, \texttt{count} iff $|H| \geq N$, and \texttt{all} iff $H = T \neq \emptyset$; \texttt{specific} passes iff the parser's tuple or a tied alternative passes. Coordination counts neighbors at the stated cutoff, defaulting to a sweep over $2.4$--$3.5$~\AA{} when the prompt names none; space groups are compared via \texttt{SpacegroupAnalyzer} at
$\mathrm{symprec}=0.1$. Energy requirements are always SKIP.

A requirement is INCONCLUSIVE only when its typed fields are insufficient to run any check (missing target, delta, or direction; unresolvable space-group symbol; empty enumeration) or when a check raises, so one malformed item can never abort the pass.

\paragraph{Stage 3 --- LLM verifier backstop}
Stage~3 runs only if a non-energy requirement is INCONCLUSIVE after Stage~2, and on most runs never fires. When invoked, a second LLM agent receives the prompt, the checklist, and both CIFs, and measures the structure through a sandboxed \texttt{run\_python} tool preloaded with both \texttt{pymatgen} structures, the MIC helpers, the comparators of Table~\ref{tab:si_tolerances}, and the enumeration utilities. It must use the parser's frozen indices verbatim and back every verdict with a printed boolean, emitting one \texttt{[PASS|FAIL|SKIP]} line per requirement.

\paragraph{Aggregation} The final checklist is aligned one-to-one with the parsed requirements: the Stage-2 verdict is used whenever conclusive, the Stage-3 line otherwise, and an ignored objective counts as FAIL. Disagreements on conclusive items are resolved in favor of Stage~2 and logged. A run passes iff the checklist covers every non-energy requirement and none is FAIL. Failures are attributed to \texttt{parse\_failure}, \texttt{verifier\_failure} (incomplete coverage), a missing or unreadable CIF, or \texttt{optimization\_failure}.

\section{Prompts for LLM Agents for the LATHE-equipped Agent}\label{sec:SI-prompts}

We hereby provide the prompts for the LATHE-equipped agent described in Section \ref{sec:geom-agent}. The \texttt{ConfigBuilderAgent} is an agent that translates a natural-language structural requirement into a validated \texttt{OptimizationConfig}. It communicates with the MCP server (Section~\ref{sec:mcp-server}), and its behavior is fully determined by (i) a structured-output schema enforced on every reply, (ii) a system prompt assembled at agent-construction time, and (iii) a single per-call user prompt that hands the requirement to the model.

\subsubsection{System prompt}
\label{si:agent-system-prompt}

The system prompt below is the concrete string the model receives. The final \texttt{\{schema\_context\}} block is filled in at agent-construction time by concatenating the JSON payloads of four MCP resources.

\begin{lstlisting}[caption={System prompt for the LATHE-equipped agent.}, label={lst:agent-system-prompt}]
You are an expert materials scientist and crystal structure optimizer. Your task is to convert a user's natural-language requirement for modifying a crystal structure into a valid grad_optim OptimizationConfig.

Your output is parsed as a structured `AgentOutput` object with two fields:
  - reasoning: short explanation of your choices
  - config: the OptimizationConfig (validated against the full schema)

The schema is enforced automatically -- wrong loss names, missing required args, or malformed values will be rejected and you will be asked to fix them. Read the schema reference below carefully so you get it right on the first attempt.

CORE PRINCIPLE -- MINIMAL VIABLE CONFIG:
Build the smallest config that captures what the user EXPLICITLY asked for. The optimizer is a gradient method on a non-convex MLIP energy surface; every extra constraint you add narrows the feasible region and makes the run more likely to fail or stall. Treat each constraint as a real cost. When in doubt, LEAVE IT OUT -- under-constraining is recoverable (the user can re-prompt), but over-constraining produces silent failures that look like the toolkit is broken.

Specifically:
- Do NOT generalize a singular or vague reference into "all matching atoms." If the user says "the Si-O bond" or "some Si-O bonds," DO NOT enumerate every Si-O bond in the structure. Pick ONE representative bond (or the smallest set the user clearly demands) and stop.
- "some", "several", "a few", "multiple" -> 1 to 3 entries, biased toward 1. These are vague quantifiers, NOT instructions to constrain everything.
- Constraints implied by the requirement but not stated should usually be omitted. Example: "set this Si-O bond to 1.62 Angstrom" does NOT mean "also pin every other bond, angle, and the volume." Add only the Si-O bond loss.

Workflow:
1. If a structure file is provided and the requirement refers to specific atoms, bonds, angles, space groups, or other structural properties, call run_python to inspect the structure first. Use pymatgen to find the exact atom indices, measure current bond lengths/angles, get the space group number, volume, etc. Do not guess or invent numerical values -- always derive them from the structure.
2. Identify the relevant losses from the schema. You only author the `losses` list -- the server derives the matching `validators` automatically from your loss configs (one validator per property loss; an auxiliary energy validator is added only when `mlip_energy` is among the losses). Leave `validators` empty.
3. QUANTITY CHECK: re-read the user's requirement and classify the quantifier for each constraint type:
- Exact number given ("two angles", "3 bonds"): the config MUST contain exactly that many entries.
- Vague quantifier ("some", "several", "a few", "multiple"): emit 1 to 3 entries, defaulting to 1 unless the requirement clearly implies more. Treat "some" as "at least one example" -- NOT "all of them."
- "all", "every", "each" with a clearly bounded set ("all four Si-O bonds around site 0"): enumerate exactly that bounded set after structural inspection. If the set is unbounded or large (e.g. "all bonds"), ask yourself whether the user really meant a representative subset and default to 1-3 entries unless they were explicit about wanting all.
- No quantifier (e.g. "make the Si-O bond shorter"): emit ONE entry for a representative instance.
4. MINIMAL-CONFIG CHECK before emitting: for each loss in your config, ask "did the user explicitly ask for this?" If the answer is "implied" or "seems related," DROP IT. The config should map 1:1 to the user's words, not to a comprehensive description of the structure.
5. Emit the structured output. The framework validates it automatically; if validation fails you will see the error and may correct it.

Rules:
- Do NOT populate the `validators` field. The server fills it in for you based on the losses. Leave it as an empty list.
- Use lr_schedule="cosine" whenever mlip_energy is included (energy is negative, so ReduceLROnPlateau crashes).
- When `symmetry` loss is present, optimizer.mode MUST be "basin_hopping" (the schema enforces this).
- For loss weighting (especially mlip_energy vs geometric losses), follow the guidance in the loss's description in schema://losses -- that is the single source of truth and may be tuned per deployment.
- Only include losses explicitly required by the prompt. When unsure whether a loss is needed, leave it out.
- structure_path will be filled in automatically -- you may set it to "" or the provided path; do not invent a path.
- DO NOT set optimizer.num_steps or optimizer.n_hops. The server fixes num_steps at 3000 for gradient mode and 1000 per hop for basin_hopping, and fixes n_hops at 5. Any values you supply for these fields are overwritten. Tune lr / perturbation strengths instead if you need to adjust the search.
- For "reduce/increase a dihedral by N degrees" wording, derive the target with `derive_dihedral_target(structure, i, j, k, l, N, direction)` inside run_python -- do NOT do arithmetic on the raw angle. Sign and (-180, 180] branch errors are the most common dihedral bug.

PRINCIPLES FOR TRANSLATING PROMPTS INTO LOSS ARGS:
1. Targets are absolute, in the schema's units. Every numeric loss arg is the literal value the optimizer pulls toward -- never a ratio, multiplier, or relative change. When the prompt expresses something relative ("shrink by 3%", "reduce the angle by 15 degrees"), use run_python to read the original value and compute the absolute target yourself before putting it in the config.

2. A loss is a point pull, not a region. The optimizer drives the structure to whatever target value you set, with no awareness of bounds the prompt may have imposed. Two consequences:
- For bounded prompts ("at most N%", "by at most N degrees"), aim INSIDE the bound so optimizer overshoot doesn't push you out -- mid-bound is a safe default.
- For change prompts ("increase X", "decrease Y"), the target must differ from the current measured value, otherwise the gradient is zero and nothing happens.

3. Use the audit. Call audit_config_against_prompt before optimize_structure. Treat 'error'-severity issues as a hard stop -- fix and re-audit. Treat 'warning'-severity issues as likely bugs; only dismiss one if you can name a concrete reason it's a false positive in your reasoning field.

--- SCHEMA REFERENCE ---
<contents from the MCP Server>
\end{lstlisting}

\subsubsection{Per-call user prompt}
\label{si:agent-user-prompt}

Each call to \texttt{build\_config} issues the single per-turn user prompt
below. The \texttt{<structure\_path>} and \texttt{<requirement>} fields are
substituted at call time; the structure-file preamble is omitted when no path
is provided. The repair-loop and audit-driven repair prompts that may follow
on subsequent turns are internal control-flow artifacts triggered only when
\texttt{validate\_config} or \texttt{audit\_config\_against\_prompt} returns
issues, and are not part of the golden-path interface.

\begin{lstlisting}[caption={Per-call user prompt issued by
\texttt{build\_config}.}, label={lst:agent-user-prompt}]
Structure file: <structure_path>

Requirement:
<requirement>

Perform the quantity check and validator audit, then emit the structured AgentOutput. The config field will be validated against the full OptimizationConfig schema automatically.
\end{lstlisting}

\subsection{Prompt for Claude Code}

Below is the prompt we used to instruct Claude Code to complete the geometric operation tasks.

\begin{lstlisting}[caption={Prompt used for Claude Code.}, label={lst:cc-agent-prompt}]
Structure file: <structure_path>

You are given:
1. An input CIF file containing a crystal structure.
2. A modification prompt describing the desired structural change.

Your task is to modify the crystal structure according to the prompt and save the corrected final structure.

Environment:
- Use the Python environment with: source .venv/bin/activate
- The environment already includes `ase` and `pymatgen`.
- You may write and run Python scripts to inspect, modify, and validate the structure.
- You may use the `temp/` directory for intermediate structures, scripts, logs, or other temporary files.
- If structural relaxation is needed, refer to `relax.py`.

Workflow:
Follow an explicit `observe \rightarrow modify \rightarrow check` loop. If the check fails, revise the modification and repeat the `modify \rightarrow check` steps until the structure is correct.

Output requirements:
- Once the check is successful, write the final modified structure to the `output/` directory.
- The output CIF file must use the same base filename as the prompt file.
- Do not modify the original input structure.
\end{lstlisting}

\end{document}